\documentclass[12pt]{article}
\input{epsf}
\def\spin{\mbox{Spin}}
\def\vol{\mbox{vol}}
\def\tfour{{\bf T}^4}
\def\R{{\bf R}}

\begin{document}

\begin{titlepage}
\begin{flushright}
NSF-ITP-00-112\\
ITEP-TH-37/00\\
\end{flushright}

\begin{center}
{\Large $ $ \\ $ $ \\
Giant Gravitons from Holomorphic Surfaces}\\
\bigskip\bigskip\bigskip
{\large Andrei Mikhailov}
\footnote{On leave from the Institute of Theoretical and 
Experimental Physics, 117259, Bol. Cheremushkinskaya, 25, 
Moscow, Russia.}\\
\bigskip
Institute for Theoretical Physics,\\
University of California, Santa Barbara, CA 93106\\
\vskip 1cm
E-mail: andrei@itp.ucsb.edu
\end{center}
\vskip 1cm
\begin{abstract}
We introduce a class of supersymmetric cycles in spacetimes
of the form AdS times a sphere or $T^{1,1}$ which
can be considered as generalizations of the giant gravitons.
Branes wrapped on these cycles 
preserve $1\over 2$, $1\over 4$ or $1\over 8$ of the supersymmetry.
On the CFT side these configurations correspond to superpositions
of the large number of BPS states. 
\end{abstract}
\end{titlepage}

\section{Introduction}
Supergravity on $AdS$ spaces has received a lot of attention
in the last few years
because of its conjectured relation to the Conformal Field Theory
\cite{Maldacena}.
This relation implies the correspondence between the primary operators 
in the bulk and the supergravity fields propagating in the $AdS$
space. The standard way to formulate this correspondence goes
as follows \cite{GKP,Witten}. 
Suppose that the Yang-Mills coupling constant $g^2_{YM}$ is very small
and the t'Hooft coupling constant $g_{YM}^2N$ is very large, 
then the dual string theory
on $AdS_5\times S^5$ should be well described by the classical supergravity.
Upon the reduction of the massless fields in the classical supergravity
on $S^5$ we get the series of massive fields in AdS, which correspond
to the primary operators in the conformal field theory.
The spectrum of the Kaluza-Klein excitations in AdS supergravities
was computed in \cite{KRN,Nieuwenhuizen,BCERS,CAFP}, and this
gives us the description of the spectrum of primary operators 
on the CFT side.
But this description is good only for those operators which have
relatively small conformal dimensions. Increasing the conformal
dimension corresponds on the supergravity side to increasing the
momentum of the Kaluza-Klein harmonic. When the momentum is high
enough, it could happen that the linearized approximation
to supergravity breaks down and we have to take into account
interactions. Let us consider for example the four-dimensional
$N=4$ supersymmetric Yang-Mills theory which is dual to the Type IIB
supergravity on $AdS_5\times S^5$. The Kaluza-Klein harmonics
are parametrized by their momentum $M$ which is an integer
(strictly speaking, a set of integers). The energy of this harmonic 
is $m={M\over R}$, where
$R$ is the radius of $AdS_5\times S^5$ (we assume that our harmonic
saturates the BPS bound). We can interpret this energy
as the mass of the corresponding five-dimensional field in $AdS_5$.
The particle with such a mass has the Schwarzschild radius
\begin{equation}
r_S\simeq\sqrt{{g_{str}^2\over R^5} m}=
{1\over m}\sqrt{M^3\over N^2}
\end{equation}
where we have taken into account $g_{str}N\simeq R^4$.
We see that for $M>>N^{2/3}$ the Schwarzschild radius
becomes larger then the Compton wavelength, which indicates
the breakdown of the linearized supergravity approximation.
Therefore the analysis of \cite{KRN,Nieuwenhuizen,BCERS,CAFP}
does not work for the high momentum states.

The supergravity description of the $1\over 2$ BPS
states with high $M$ was found by
McGreevy, Susskind and Toumbas \cite{MGST}. 
They have found that these states should be considered
not as  Kaluza-Klein reductions of massless supergravity fields
but as branes. 
This is an example of  a general phenomenon:  the massless particle 
can ``polarize'' and become a brane \cite{Myers}.
The branes representing the high momentum states were called 
giant gravitons.

Giant gravitons preserving half of the supersymmetries have been
studied in \cite{MGST,GMT,HHI}. It has been found that they
come in two species.  The first type are spheres inside $S^5$,
the projection of the worldvolume on the AdS space being a timelike geodesic. 
The second type are the spheres inside $AdS_5$, the projection
on a sphere being an equator. (These two types are formally related
by an analytic continuation, which relates $AdS_5\times S^5$ to
$S^5\times AdS_5$.) In our paper we discuss more general configurations 
preserving $1\over 4$ or $1\over 8$ supersymmetry. 
We present a class of such configurations 
parametrized by the holomorphic curves in ${\bf C}^3$.
We formulate our results and explain the CFT meaning of our
configurations in Section 2. In Sections 3,4 and 5 we prove 
that our configurations are supersymmetric. 
In Section 6 we derive the mass formula and verify the saturation of 
the BPS bound. 
In Section 7 we consider similar configurations in flat space. 
In Appendix we comment on the relation between our results 
and the results of the linearized approximation which was developed in 
\cite{DJM}.

\section{The construction}
\subsection{$AdS_5\times S^5$ and $AdS_5\times T^{1,1}$}
Consider a five-dimensional manifold $M$ embedded into a six-dimensional 
$N$. In our application, $M$ is a part of the $1+9$-dimensional space-time 
of Type IIB string theory, and $N$ is an auxiliary manifold 
(the orthogonal direction to $M$ in $N$ does not belong to the space-time 
and does not have any physical meaning). For example, we can take $M=S^5$ 
of $AdS_5\times S^5$, and $N$ to be ${\bf R}^6$. Suppose that $N$ 
has a complex structure. In this situation, $M$ has a preferred field 
of directions. This field of directions is defined as follows. 
Let $e^{\perp}$ be the unit vector in $TN$ orthogonal to $TM$.
Then, $e^{||}=I.e^{\perp}$ belongs to the tangent space of $M$.
This gives us the field of directions in $M$.

\underline{
Given a holomorphic surface $C\subset N$, we associate to it a giant
graviton.} We will do it in the following way.
Let us denote $\Sigma$ the intersection of $C$ with $M$.
We will consider $\Sigma$ to be the surface of the $D3$ brane at
time $\tau=0$. Let us move this brane by translation in the 
preferred direction $e^{||}$ with the speed of light.
Since $e^{||}$ is usually not orthogonal to $\Sigma$, 
the surface elements of the brane actually move with the
speed less than the speed of light. We claim that the resulting
$1+3$-dimensional worldvolume is supersymmetric.

\underline{Example.} Let us consider $S^5\subset {\bf C}^3$.
We denote $Z_1,Z_2,Z_3$ the flat coordinates on ${\bf C}^3$,
and $S^5$ is $|Z_1|^2+|Z_2|^2+|Z_3|^2=1$. Suppose that $C$
is given by the equation $Z_1=\rho$. Then, $\Sigma=S^3$,
and the resulting worldvolume is the giant graviton 
found in \cite{MGST}.
The spatial trajectory of this giant graviton is $S^1\times S^3$.

More generally, we can consider $C$ described by 
the equation of the form $F(Z_1,Z_2,Z_3)=0$, and the worldsurface
of the giant graviton at time zero will be 
\begin{equation}
\left\{ \begin{array}{l}
|Z_1|^2+|Z_2|^2+|Z_3|^2=1\\
F(Z_1,Z_2,Z_3)=0
\end{array}
\right.
\end{equation}
Suppose that we measure time as the length of the geodesic
in AdS. Then, the worldsurface of the giant graviton at time $t$
will be:
\begin{equation}
\left\{ \begin{array}{l}
|Z_1|^2+|Z_2|^2+|Z_3|^2=1\\
F(e^{-it/R}Z_1,e^{-it/R}Z_2,e^{-it/R}Z_3)=0
\end{array}
\right.
\end{equation}
where $R$ is the radius of $S^5$. This is a supersymmetric trajectory.

We can apply our construction for gravitons in $AdS_5\times T^{1,1}$.
In this case, we should embed $T^{1,1}$ into the conifold, as explained
in \cite{CdlO}:
\begin{equation}
\begin{array}{c}
X_1^2+\ldots+X_4^2=0,\\
|X_1|^2+\ldots+|X_4|^2=1
\end{array}
\end{equation}
The manifold $T^{1,1}$ is a $U(1)$ bundle over $S^2\times S^2$,
and the preffered direction $e^{||}$ is in fact the direction along the
fiber. Translation along this preffered direction corresponds to the
phase rotation of $X_I$. The equation for the giant graviton is:
\begin{equation}
\left\{
\begin{array}{l}
X_1^2+\ldots+X_4^2=0,\\
|X_1|^2+\ldots+|X_4|^2=1,\\
F(e^{-it/R}X_1,\ldots,e^{-it/R}X_4)=0
\end{array}
\right.
\end{equation}
We have not tried to turn on the gauge field living on the $D3$ brane.

\subsection{$AdS_3\times S^3$}
The spacetime $AdS_3$ arises in the study of D-branes of Type IIB 
string theory on ${\bf T}^4$ or $K3$. We will consider
Type IIB on ${\bf T}^4$. D-branes with $1+1$
noncompact dimensions are effectively described as
six-dimensional black strings. These black strings are classified
by the charge $Z\in H^*(\tfour,{\bf Z})$. Take a large number
$N$ of black strings all having the same charge $Z$. The geometry of
the near horizon region is $AdS_3\times S^3\times {\bf T}^4$.
Consider the probe string having the same charge $Z$.
Suppose that at the time $t=0$ the probe string spans a one-dimensional
curve $\gamma\subset S^3$, and is a point in $AdS_3$.
Let us think of $S^3$ as a group manifold of $SU(2)$. Suppose
that our probe string moves along a geodesic line in $AdS_3$
and its position in $S^3$ changes as follows:
\begin{equation}
\gamma\mapsto e^{it\sum_{i=1}^3 x_i\sigma^i}.\gamma
\end{equation}
where $\sigma^i$ are Pauli matrices, $\sum_{i=1}^3 x_i^2=1$,
$t$ is the length parameter along the geodesic.
In other words, we translate our effective string along the one-parameter
subgroup in $SU(2)$ with the speed of light. 
We claim that this brane configuration is supersymmetric, preserving
$1\over 4$ of the supersymmetries. We will prove it in Section 4.

This gives us a class of supersymmetric configurations 
in $AdS_3\times S^3$ parametrized by the one-dimensional curves
$\gamma\subset S^3$.

\subsection{Giant gravitons in $AdS_4\times S^7$ and $AdS_7\times S^4$}
The construction of giant gravitons for $AdS_4\times S^7$ is similar
to the construction for $AdS_5\times S^5$. We embed $S^7$ into
${\bf C}^4$ and consider intersections with the holomorphic surfaces
in ${\bf C}^4$. We get configurations preserving $1\over 2$, 
$1\over 4$ and $1\over 8$ supersymmetries.

Supersymmetric cycles in $AdS_7\times S^4$ can be constructed in the
following way. We embed $S^4\subset {\bf R}^5$ and represent
${\bf R}^5={\bf R}\times {\bf C}^2$. Then we consider the cylindrical
surfaces in ${\bf R}^5$ of the form ${\bf R}\times C$ where 
$C\subset {\bf C}^2$ is a holomorphic curve. Intersection of these
cylinders with $S^4$ gives us the worldvolumes of $M2$ branes at
time zero. The motion of the $M2$ corresponds to multiplication
of the coordinates of ${\bf C}^2$ by $e^{-it}$.
These configurations for generic $C$ 
preserve $1\over 4$ of the supersymmetries. 
If $C$ is a plane 
then we reproduce the giant graviton of \cite{MGST} which preserves
$1\over 2$ of the supersymmetry.
We can interpret the splitting ${\bf R}^5={\bf R}\times {\bf C}^2$
in the following way. The momentum of the BPS state is an element
of $so(5)$ and it can be represented as an antisymmetric matrix $M_{IJ}$
acting in ${\bf R}^5$. An antisymmetric matrix acting in
an odd-dimensional space has at least one zero eigenvector.
The preffered direction ${\bf R}$ is generated
by a vector annihilated by $M_{IJ}$.

The details for $AdS_4\times S^7$ and $AdS_7\times S^4$ are presented
in Section 5. 

\subsection{Giant Gravitons and BPS States}
We have seen that there is a continuum of supersymmetric giant gravitons
with the given momentum, parametrized by the holomorphic
curves (or just arbitrary curves, in the case of $AdS_3\times S^3$).
One could think that this implies that there is a continuum of BPS states with
a given momentum on the CFT side.  Of course, this conclusion is wrong.
To find the degeneracy of the BPS states one should study the giant gravitons
quantum mechanically. Classical supersymmetric solutions of the D-brane
equations of motion corresponding to the giant gravitons are in fact
superpositions of the large number of supersymmetric quantum states having
nearly the same energy and momentum. Let us think of the momentum $M$ as
specifying an irreducible representation of the $R$-symmetry group 
(essentially, a set of integers). What is the number of states
with the given momentum $M$? On the CFT side, we 
at least naively expect it to depend
on $M$ and the number of colors $N$: ${\cal N}={\cal N}(M,N)$ ---
it is the number of primary operators transforming in a given
representation of the R-symmetry group.
On the supergravity side $N$ is proportional $TR^{p+1}$ where
$T$ is the tension of the probe brane (the brane which makes
the giant graviton) and $R$ is the radius of the AdS space. The quantity
$TR^{p+1}$ is in fact the only dimensionless parameter in the
Born-Infeld action for the probe brane, if we forget about 
$\alpha'$-corrections.  Naively this implies
that the number of states with the given momentum is a function
of $TR^{p+1}$, which would agree with what we expect from CFT.
But one should remember that the Born-Infeld action is 
not enough to define a worldvolume theory on a D-brane. There are 
divergences which should be regularized by adding $\alpha'$
corrections. In fact, considering the full quantum theory on the
D brane requires taking into account the degrees of freedom 
living in the bulk.
Either the degeneracy of the BPS states of the D brane does not receive
$\alpha'$ corrections, or the number of primary operators in CFT
depends on the coupling constant. 

We will not try to quantize the worldvolume theory of the giant
graviton in our paper. 
But we want to consider a simple and well-known example in flat space 
where we can actually compute the degeneracy. 

\underline{Example.} Classical BPS configurations of the D-string.

Consider Type IIB at weak coupling $g_{str}\to 0$ 
compactified on a circle $S^1$ of the radius $R_9$.
Let us wrap $D1$ on $S^1$. Let $X^0$ be the time, $X^9$ the coordinate
on the circle, and
$X^I(\tau,\sigma)$, $I=1,\ldots,8$ the coordinates in the noncompact
directions. Consider the following classical solution of
the D-string equations of motion:
\begin{equation}\label{LeftMoving}
X^I(\tau,\sigma)=f^I(\sigma+\tau)
\end{equation}
This is the left moving wave.
This solution preserves $1\over 4$ of the supersymmetry of type IIB
(we review this fact in Section 7).
It saturates the BPS bound for a given momentum.
Indeed, the energy of the solution is equal to:
\begin{equation}
E=T\int dl {1\over\sqrt{1-v_{\perp}^2}}
 =T\int {dl^2\over dX^9}=2\pi R_9T+T\int dX^9 \sum_{I=1}^8 
\left({dX^I\over dX^9}\right)^2
\end{equation}
where $T$ is the tension of the D-string.
We have taken into account that $v_{\perp}^9=\left({dX^9\over dl}
\right)^2-1$, $v_{\perp}^I={dX^9\over dl}{dX^I\over dl}$
and $(v_{\perp})^2=1-\left({dX^9\over dl}\right)^2=
\sum_{I=1}^8\left({dX^I\over dl}\right)^2$.
The momentum along the circle is:
\begin{equation}
P_9=T\int dl {v_{\perp}^2\over\sqrt{1-v_{\perp}^2}}=
T\int dX^9\sum_{I=1}^8 \left({dX^I\over dX^9}\right)^2
\end{equation}
We see that the BPS bound is saturated:
\begin{equation}
E=2\pi R_9 T+P_9
\end{equation}
Quantum mechanically $P_9$ should be quantized and for each integer 
$P_9$ there is a finite number of states. Let us briefly review how this
happens.
Suppose that the coordinates are slowly varying on the worldsheet:
\begin{equation}
\begin{array}{c}
S \simeq
-2\pi R T\int dX^0\left[ 1 -{1\over 4\pi R}\int dX^9\left( 
\left|\left|{\partial X^I\over \partial X^0}\right|\right|^2
-\left|\left|{\partial X^I\over \partial X^9}\right|\right|^2
\right)\right]
\end{array}
\end{equation}
We will forget about fermions.
Consider the Fourier decomposition of the worldsheet fields:
\begin{equation}
X^I(X^0,X^1)=\sum_{k=-\infty}^{\infty}X^I_k(X^0)e^{ikX^9/R}
\end{equation}
The action can be rewritten in terms of the Fourier modes:
\begin{equation}
S=
-2\pi R T\int dX^0\left[ 1 -{1\over 2}\sum_{k=-\infty}^{\infty}
\left( 
{dX_{-k}^I\over dX^0}{dX_k^I\over dX^0}
-{k^2\over R^2} X^I_{-k}X^I_k
\right)\right]
\end{equation}
The Hamiltonian is:
\begin{equation}
H={1\over R} \sum_{k>0}\left(||a^{kI}||^2
+||\tilde{a}^{kI}||^2\right)+\mbox{const}
\end{equation}
where
\begin{equation}
\begin{array}{c}
a^{kI}={1\over\sqrt{2}}
\left({1\over\sqrt{2\pi T}}P^{kI}+ik\sqrt{2\pi T}X^{-kI}\right)\\
\tilde{a}^{kI}={1\over\sqrt{2}}
\left({1\over\sqrt{2\pi T}}P^{kI}-ik\sqrt{2\pi T}X^{-kI}\right)
\end{array}
\end{equation}
and $P^{kI}$ is the momentum conjugate to $X_k^I$,
$P^{Ik}=2\pi RT{dX_{-k}^I\over dX^0}$.
In quantum mechanics $a$ and $\tilde{a}$ become creation/annihilation 
operators, ${ [ }a^{kI},a^{-k'I'}{ ] }=-k\delta_{k,k'}\delta_{I,I'}$.
Let us consider the states of the form:
\begin{equation}
(a^{k_1I_1})^{n_1}\cdots (a^{k_pI_p})^{n_p}|0>
\end{equation}
These states are annihilated by $\tilde{a}^{kI}$ with negative
$k$. Quasiclassically this means that the state is purely left moving.
These are the BPS states. The classical BPS configurations given
by (\ref{LeftMoving}) are superpositions of the large numbers of
these quantum BPS states.

Suppose that we start increasing the string coupling constant.
When $g_{str}\to\infty$ we can describe our D-string as a fundamental
string of the dual Type IIB. In this limit the left-moving
waves correspond to the BPS states in the spectrum of the fundamental
string which have been discovered by Dabholkar and Harvey
\cite{DH}. These states are the form
\begin{equation}
\begin{array}{l}
(\alpha_{-k_1}^{I_1})^{n_1}\cdots 
(\alpha_{-k_p}^{I_p})^{n_p}|w_9,p_9>
\\
\sum k_pn_p=w_9p_9
\end{array}
\end{equation}
where $p_9$ is related to the momentum $P_9={1\over R}p_9$ 
and $w_9$ is the winding number;
in our case $w_9=1$. The degeneracy of these states is
${\cal N}(p_9)$ --- the coefficient of $q^{p_9}$ in 
$\prod_{m=1}^{\infty}{1\over (1-q^m)^8}$ (in fact there is more degeneracy
because of the worldsheet fermions).
The number of states with the given momentum $P_9$ is finite although rapidly
growing with $P_9$. 

In this example, the degeneracy of states with the given momentum did not
depend on $TR^2$. This is because the theory on the worldvolume
was free. If we have added interactions, for
example consider the D string living in the curved space
then the spectrum would depend on $Tr^2$ where $r$ determines
the curvature. If in addition these interactions
were nonrenormalizable then the spectrum would also depend on the
cutoff (that is on $\alpha'$).

\section{Supersymmetry for $1\over 8$ BPS giant gravitons
\\ in $AdS_5\times S^5$}
\subsection{Spinors in $AdS_{p-1}\times S^{q-1}$}
Spinors in the spaces of the form AdS times a sphere
were studied in many papers \cite{Bar,KW,Imamura,Kehagias}.
In this subsection we will briefly review this subject.

First of all, let us remember what is the spin bundle. 
Consider a manifold $X$ with a metric $g_{\mu\nu}(x)$.
Let us choose a field of tetrads $e^1(x),\ldots,e^N(x)$,
where $N=\mbox{dim}X$ and $g_{\mu\nu}=e^m_{\mu}e^m_{\nu}$, 
summation over $m$. A section of the spinor bundle $S(TX)$
is defined with respect to the choice of the field of tetrads;
it is a field $\psi(x)\in {\cal S}$  where ${\cal S}$ is a spinor
representation of $SO(N)$. When we change the field
of tetrads $\psi(x)$ should change to $\rho_s(g(x))\psi(x)$,
where $g(x)$ is the unique element of $SO(N)$ such that
the new tetrad is $e'(x)=g(x)e(x)$ and $\rho_s(g(x))$ is the
spinor representation of $g(x)$. There is a sign ambiguity in the
choice of the spin representation, but it is not very important 
for us now. $S(TX)$ comes equipped with the connection, which is
defined as follows.
The section of the spinor bundle is covariantly constant 
along the path $x(\tau)$ if and only if $\psi(x)$ is constant provided 
that we have chosen the field of tetrads in such a way that 
${D\over D\tau}e^m(x(\tau))=0$. This connection is called 
``spin connection''. It can be described in the following way.
Let us choose $N$ {\em constant} gamma-matrices $\Gamma^m$,
satisfying $\Gamma^m\Gamma^n+\Gamma^n\Gamma^m=2\delta^{mn}$.
Then, 
\begin{equation}
\nabla_{\mu}\psi(x)=\left(\partial_{\mu}+
{1\over 4}\omega^{mn}_{\mu}\left[\Gamma^m,\Gamma^n\right]
\right)\psi
\end{equation}
where $\omega^{mn}$ is a one-form determined from 
$de^m=\omega^{mn}\wedge e^n$.

Suppose that we have an embedding of manifolds $i:X\subset Y$, 
$\mbox{dim}X=k$, $\mbox{dim}Y-\mbox{dim}X=l$, $l$ is even,
both $X$ and $Y$ have a spin structure.
The structure group of $i^*(TY)$ is $\spin(k)\times\spin(l)$.
In other words, we can define the field of tetrads in such a way that
the first $k$ vectors $e_m$ belong to $TX$ and the last
$l$ of them belong to $(TX)^{\perp}$.
This gives us an isomorphism
\begin{equation}
R: \;\;
S(TX)\otimes S((TX)^{\perp})\widetilde{\rightarrow}
i^*S(TY)
\end{equation}
However, one should remember that this isomorphism usually does
not commute with the parallel transport.

It is important for us to consider a special case with $Y=N_1\times N_2$,
$X=M_1\times M_2$, $M_i\subset N_i$, $\mbox{dim}N_i-\mbox{dim}M_i=1$.
For example, we can take $N={\bf R}^{2+4}\times {\bf R}^6$ and
$M=AdS_5\times S^5$. More generally, we assume that $N_i$ is the cone
with the base $M_i$. This means that the metric on $N_i$ can be written
as follows:
\begin{equation}
ds_{M_i}^2=\left({dR\over R}\right)^2+e^{2f(R)}ds_{N_i}^2
\end{equation}
We can write a general expression for some components of the spin 
connection:
\begin{equation}
\omega^{iR}=-{\partial f\over\partial R}e^i
\end{equation}
where $e^i$ is tangent to $M_i$.
Suppose that $\Psi_{++}$ is a covariantly constant spinor in $N_1\times N_2$ 
satisfying $\Gamma^{1\ldots p}\Psi_{++}=-i^{p\over 2}\Psi$
and $\Gamma^{p+1\ldots p+q}\Psi_{++}=-i^{q\over 2}\Psi$. 
Then, the restriction
of the covariantly constant spinor to $M_1\times M_2$ satisfies the
following equations:
\begin{equation}\label{CovariantlyConstant}
\begin{array}{c}
\left({D\over D x^{\mu}}-i^{p\over 2}
{\partial f\over\partial R}\Gamma^{\mu}\prod\limits_{I=1}^{p-1}\Gamma^I
\right)\Psi_{++}=0\;\;\; (\mu=1,\ldots,p-1)
\\
\left({D\over D y^{\mu}}-i^{q\over 2}
{\partial f\over\partial R}\Gamma^{p+\mu}\prod\limits_{I=1}^{q-1}
\Gamma^{p+I}
\right)\Psi_{++}=0\;\;\; (\mu=1,\ldots,q-1)
\end{array}
\end{equation}
where $x^{\mu}$ are the coordinates on $M_1$ and $y^{\mu}$ are
the coordinates on $M_2$.
These are the equations for covariantly constant spinors in 
$AdS_5\times S^5$. 
$\Psi_{++}$ is not the only solution to the equations
(\ref{CovariantlyConstant}). One can see that 
$\Gamma^p\Gamma^{p+q}\Psi_{++}$
is another solution. In fact, our spinor representation is reducible 
if considered as a representation of $\Gamma^1,\ldots,\Gamma^{p-1},
\Gamma^{p+1},\ldots,\Gamma^{p+q-1}$. The operator 
${1\over 2}(1+i\Gamma^p\Gamma^{p+q})$ is the projector on the irreducible 
subspace. This gives us the following description of the Killing spinors
in $AdS_{p-1}\times S^{q-1}$:
\begin{equation}
\Psi={1\over 2}(1+i\Gamma^p\Gamma^{p+q})\Psi_{++}
\end{equation}
Here we have assumed that both the AdS and the sphere have Euclidean
signature. The formula for the Minkowski AdS times the Euclidean sphere
is the same except for one should skip $i$.

\subsection{Some geometric background}
First of all, let us summarize
already introduced notations and introduce some more.
We have embedded $M\subset N$, $\mbox{dim}\;N-\mbox{dim}\;M=1$,
$N$ has a complex structure $I$, which is an operator in $TN$ with
$I^2=-1$ satisfying certain integrability conditions. For our 
application, we can take $M=S^5$ and $N={\bf C}^3$, or
$M=T^{1,1}$ and $N=$conifold. We are considering a holomorphic
surface $C\subset N$, $\mbox{dim}_{\bf C} C=\mbox{dim}_{\bf C}N-1$.
In this situation, it is natural to consider the following obects:

\begin{tabular}{|c|l|}
\hline
$e^{\perp}$ & 
Unit vector orthogonal to $TM\subset TN$\\
\hline
$e^{||}$ & $I.e^{\perp}$ \\
\hline
$\Sigma$ & $C\cap M$ \\
\hline
$(TC)^{\perp}$ & Orthogonal complement of $TC$ in $TN$\\
\hline
$e^{\phi}$ & Unit vector in $(TC)^{\perp}\cap TM$; notice that
$e^{\phi}\perp T\Sigma$ \\
\hline
$T_0\Sigma$ & Maximal subspace in $T\Sigma$ closed
under $I$, $I.T_0\Sigma=T_0\Sigma$\\
\hline
$e^{\psi}$ & Unit vector  in the orthogonal complement to $T_0\Sigma$
in $T\Sigma$\\
\hline
$e^n$ & Unit vector in $TM$ orthogonal to $e^{\phi}$ and $T\Sigma$\\
\hline
\end{tabular}

Notice that the orthogonal complement to $T\Sigma$ in $TM$ has real dimension
two. In this subsection we will prove that the component of $e^{||}$
orthogonal to $T\Sigma$ is proportional to $e^{\phi}$:
\begin{equation}
e^{||}=\lambda e^{\phi}\;\;\mbox{mod}\; T\Sigma,\;\;\lambda\leq 1
\end{equation}
We will also find that the component of $e^{||}$ parallel to $T\Sigma$
is proportional to $e^{\psi}$ (in other words, it is orthogonal
to $T_0\Sigma$).

Consider $I.e^{\phi}$. This vector should be
orthogonal to $e^{\phi}$, and also orthogonal to $\Sigma$
(because $(TC)^{\perp}$ is closed under $I$).
This means that the three vectors  $I.e^{\phi}$,
$e^{\perp}$ and $e^n$ are linearly dependent: 
\begin{equation}\label{LinearDependence}
I.e^{\phi}=\cos\alpha e^{\perp}+
\sin\alpha e^n
\end{equation}
Taking the scalar product of this equation with $e^{||}$, we find that
$e^{||}$ is orthogonal to $e^n$. This proves our first statement:
the component of $e^{||}$ orthogonal to $T\Sigma$ is directed
along $e^{\phi}$. Suppose that $v\in T_0\Sigma$. Then, 
$(e^{||}\cdot v)=-(e^{\perp}\cdot I.v)=0$ because $I.v$ also
belongs to $T_0\Sigma$. Therefore, $e^{||}$ is orthogonal to $T_0\Sigma$,
which is our second statement. Now, we have $e^{||}$ a linear combination
of $e^{\phi}$ and $e^{\psi}$:
\begin{equation}\label{ExpressionForPrefferedE}
e^{||}=-\cos\beta e^{\phi}+\sin\beta e^{\psi}
\end{equation}
Applying $I$ to (\ref{LinearDependence}) we see that $I.e^n$ is also a linear 
combination of $e^{\phi}$ and $e^{\psi}$. The coefficients are partially fixed
by $(I.e^n\cdot e^{||})=0$. 
\begin{equation}
I.e^n=\mp(\sin\beta e^{\phi}+\cos\beta e^{\psi})
\end{equation}
Comparing this with (\ref{LinearDependence}) we find $\beta=\pm \alpha$.

We have proven the following statement: 
the normal component of $e^{||}$ is $-\cos\alpha e^{\phi}$,
where $\cos\alpha$ is the coefficient in the linear dependence
(\ref{LinearDependence}).

\subsection{Supersymmetry for $1\over 8$ BPS giant gravitons}
We will use an embedding
$$
AdS_5\times S^5\subset {\bf R}^{2+4}\times {\bf R}^{6}
$$
We enumerate gamma-matrices in such a way that $\Gamma^5$
corresponds to the direction orthogonal to $AdS_5$ in ${\bf R}^{2+4}$
and $\Gamma^{11}$ corresponds to the direction in ${\bf R}^{6}$
orthogonal to $S^5$. We will concentrate on the graviton spread
inside $S^5$. We will think of $S^5$ as $M$ and ${\bf R}^6={\bf C}^3$
as $N$. Given a vector $v$, we denote $\Gamma(v)$ the corresponding
gamma-matrix. Combining these notations with the notations
from the previous paragraph we can write
$\Gamma^{11}=\Gamma(e^{\perp})$.
Our sign conventions imply $(\Gamma^0)^2=(\Gamma^5)^2=1$
and the other $\Gamma^I$ have $(\Gamma^I)^2=-1$.
The covariantly constant spinor in $AdS_5\times S^5$ has a form
\begin{equation}
\Psi=(1+\Gamma^5\Gamma^{11})\Psi_{++}
\end{equation}
where $\Psi_{++}$ is a constant spinor subject to the following constraints:
\begin{equation}\label{PlusPlus}
\begin{array}{c}
\Gamma^0\cdots\Gamma^5\Psi_{++}=i\Psi_{++}\\
\Gamma^6\cdots\Gamma^{11}\Psi_{++}=i\Psi_{++}
\end{array}
\end{equation}
Let $\Gamma^{\tau}$ corresponds to the direction of the timelike
geodesic in $AdS_5$. The condition for preserved supersymmetry is
\begin{equation}
{1\over\sqrt{1-v^2}}(\Gamma^{\tau}+v\Gamma^{\phi})\hat{\Sigma}\Psi=i\Psi
\end{equation}
Here $\Psi=\Psi_L+i\Psi_R$ is a combination of the Majorana-Weyl 
generators of supersymmetry in Type IIB. 
$\hat{\Sigma}={1\over 6}\Sigma_{\alpha\beta\gamma}\Gamma^{\alpha}
\Gamma^{\beta}\Gamma^{\gamma}$ is the gamma-matrix representation of
the surface element of the $D3$ brane. $\Gamma^{\phi}=\Gamma(e^{\phi})$
is the gamma-matrix corresponding to the vector $e^{\phi}$
introduced in the previous section. Let us parametrize
the velocity in terms of the angle $\nu$: $v=\cos\nu$.
We can rewrite the supersymmetry condition as follows:
\begin{equation}\label{SUSYForDThree}
\begin{array}{c}
0=\left[i(\Gamma^{\tau}+\cos\nu\Gamma^{\phi})\hat{\Sigma}
+\sin\nu\right](1+\Gamma^5\Gamma^{11})\Psi_{++}
=\\=
(1+\Gamma^5\Gamma^{11})
\left[\Gamma^{\tau}\Gamma^5\Gamma^n\Gamma^{\phi}
-\cos\nu\Gamma^n\Gamma^{11}+\sin\nu\right]\Psi_{++}
\end{array}
\end{equation}
Let us impose the following constraint on $\Psi_{++}$:
\begin{equation}\label{ConstraintOnPsi}
\Gamma^{\tau}\Gamma^5\Gamma(e^{\phi})\Gamma(Ie^{\phi})
\Psi_{++}=\Psi_{++}
\end{equation}
Now we can see that the supersymmetry condition is satisfied
for $\nu=-\alpha$:
\begin{equation}
(1+\Gamma^5\Gamma^{11})
\left[\Gamma^n\Gamma(Ie^{\phi})
-\cos\alpha\Gamma^n\Gamma^{11}-\sin\alpha\right]\Psi_{++}=0
\end{equation}
which is a consequence of (\ref{LinearDependence}).

The constraint (\ref{ConstraintOnPsi})
is obviously compatible with (\ref{PlusPlus}).
We can construct a spinor satisfying this constraint in the following
way. Let us consider $\Psi_{++}$ annihilated by all the gamma-matrices
corresponding to the holomorphic directions in ${\bf C}^3$,
and belonging to the $+i$-eigenspace of $\Gamma^{\tau}\Gamma^5$:
\begin{equation}\label{StrongConditionsOnPsi}
\begin{array}{l}
\Gamma\left(\partial_{Z^I}\right)\Psi_{++}=0,\;\;
I=1,2,3\\
\Gamma^{\tau}\Gamma^5\Psi_{++}=i\Psi_{++}
\end{array}
\end{equation}
It makes sense to impose the constraint 
$\Gamma^{\tau}\Gamma^5\Psi_{++}=i\Psi_{++}$ because $\Gamma^{\tau}\Gamma^5$
is constant. Indeed, 
the timelike geodesics in $AdS_5$ are intersections with the
$2+0$-dimensional planes in ${\bf R}^{2+4}$ passing through the
origin. Therefore, $\Gamma^{\tau}\Gamma^5$ is the gamma-matrix
representation of the tangent bivector to such a plane, which is 
constant:
\begin{center}
\leavevmode
\epsffile{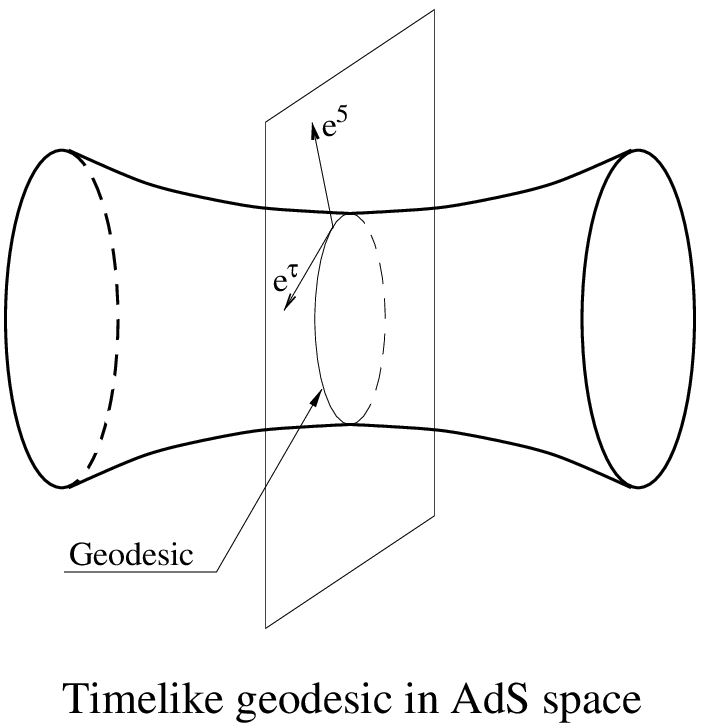}
\end{center}
Notice that the conditions (\ref{StrongConditionsOnPsi})
are generally speaking stronger then (\ref{ConstraintOnPsi}).
Only $1\over 8$ of $\Psi_{++}$ satisfy them. 
For a class of $C$, we can find more spinors satisfying 
(\ref{ConstraintOnPsi}). Suppose that $C$ is described by $F=0$
where $F=F(Z_1,Z_2)$ depends only on the first two  of the  three $Z$.
This means that $e^{\phi}$ and $I.e^{\phi}$ belong to 
${\bf C}^2\subset  {\bf C}^3$, the subset of ${\bf C}^3$ parametrized
by $Z_1$ and $Z_2$. 
In this case, besides (\ref{StrongConditionsOnPsi}), we can also
solve (\ref{ConstraintOnPsi}) as follows:
\begin{equation}
\begin{array}{l}
\Gamma\left(\partial_{\bar{Z}^1}\right)\Psi_{++}
=\Gamma\left(\partial_{\bar{Z}^2}\right)\Psi_{++}=0,\;\;
\Gamma\left(\partial_{Z^3}\right)\Psi_{++}=0
\\
\Gamma^{\tau}\Gamma^5\Psi_{++}=-i\Psi_{++}
\end{array}
\end{equation}
Therefore, in this case we will have $1\over 4$ supersymmetries
preserved.
The giant graviton of \cite{MGST} preserves
$1\over 2$ supersymmetry, and it is natural to think of it as
given by $F(Z_1)=0$, $F$ depends only on $Z_1$.

\section{Supersymmetry for $1\over 4$ BPS giant gravitons \\ in
$AdS_3\times S^3$}
Let us consider Type IIB compactified on ${\bf T}^4$,
and wrap $N$ $D3$ branes on a two-cycle $c$ in ${\bf T}^4$.
The near horizon geometry is $AdS_3\times S^3$.
Let $\alpha^{\vee}$ be a constant two-form on ${\bf T}^4$
representing the Poincare dual to $c$ and 
$\alpha_{IJ}={1\over 2}\epsilon_{IJKL}(\alpha^{\vee})^{KL}$.
Then the five-form will have the following structure:
\begin{equation}
F=\vol_{AdS_3}\wedge\alpha+\vol_{S^3}\wedge\alpha^{\vee}
\end{equation}
${\bf T}^4$ has a sphere worth of complex structures; let us
choose the one in which $c$ is represented by a holomorphic cycle.
We will denote this complex structure $I$.
Consider the embedding  
$AdS_3\times S^3\times\tfour\subset \R^{2+2}\times \R^{0+4}
\times \tfour$. We denote the coordinates in ${\bf R}^{2+2}$
$X^0,\ldots, X^3$, the coordinates in ${\bf R}^{0+4}$ 
$X^5,\ldots,X^7$, and the coordinates in ${\bf T}^4$
$X^8,\ldots, X^{11}$.
The Killing spinors on $AdS_3\times S^3$ are
parametrized by a constant spinor $\Psi_{+++}$ in ${\bf R}^{2+10}$
which satisfies the following constraints:
\begin{eqnarray}
\Gamma^8\Gamma^9\Gamma^{10}\Gamma^{11}\Psi_{+++}=\Psi_{+++}
\label{First}\\
\Gamma^0\Gamma^1\Gamma^2\Gamma^3\Psi_{+++}=
\Gamma(v)\Gamma(I.v)\Psi_{+++}\label{Second}\\
\Gamma^4\Gamma^5\Gamma^6\Gamma^7\Psi_{+++}=
i\Gamma(v)\Gamma(I.v)\Psi_{+++}
\label{Third}
\end{eqnarray}
where $v$ is an arbitrary vector in $T\tfour$ (if (\ref{Second})
and (\ref{Third}) is true for some $v$, then it is true for any 
$v$ because of (\ref{First}).) Such a $\Psi_{+++}$ is an eigenvector
of $\alpha_{IJ}\Gamma^{7+I}\Gamma^{7+J}$ as well as 
of $(\alpha^{\vee})_{IJ}\Gamma^{7+I}\Gamma^{7+J}$. 
Given $\Psi_{+++}$ satisfying
(\ref{First}-\ref{Third}), the Killing spinor is
\begin{equation}
\Psi=(1+\Gamma^3\Gamma^7)\Psi_{+++}
\end{equation}
Let us probe the near-horizon background with the brane
of the same kind: the $D3$ wrapped on $c$. The supersymmetry
condition is:
\begin{equation}\label{SUSYDOne}
\begin{array}{c}
0=\left[(\Gamma^{\tau}+\cos\nu\Gamma^{\phi})\Gamma^{\psi}
+\sin\nu\right](1+\Gamma^3\Gamma^{7})\Psi_{+++}
=\\=
(1+\Gamma^3\Gamma^7)
\left[-\Gamma^{\psi} 
(\Gamma^{\tau}+\cos\nu\Gamma^{\phi})
+\sin\nu\right]\Psi_{+++}
\end{array}
\end{equation}
where $\Gamma^{\psi}$ is the gamma-matrix corresponding to 
$e^{\psi}$ --- the unit vector tangent to the black string.
Just as in the previous section, we have 
\begin{equation}
e^{\phi}=-{1\over \cos\beta}e^{||}+{\sin\beta\over\cos\beta}e^{\psi}
\end{equation}
--- now this is just the statement that $e^{\phi}$ is the
projection of $e^{||}$ on the space orthogonal to the black string.
Let us introduce a complex structure $J$ in $\R^4$ in such a way
that 
\begin{equation}\label{EparIsJEperp}
e^{||}=J.e^{\perp}
\end{equation}
 We impose the following conditions on
$\Psi_{+++}$:
\begin{equation}\label{ConditionsDOne}
\begin{array}{c}
\Gamma^4\Gamma^5\Gamma^6\Gamma^7\Psi_{+++}=\Psi_{+++}\\
\Gamma^{\tau}\Gamma^3\Gamma^{||}\Gamma^7\Psi_{+++}=-\Psi_{+++}
\end{array}
\end{equation}
If the second of these two conditions is satisfied at some point
of $S^3$, then it is also satisfied at any other point.
This follows from the first condition and from (\ref{EparIsJEperp})
(remember that in our notations $\Gamma(e^{\perp})$ is the same as
$\Gamma^7$). 

One can see that (\ref{SUSYDOne}) follows from
(\ref{ConditionsDOne}).
Therefore, the black string translated with the speed of light
in the direction $e^{||}$ preserves $1\over 4$ of the supersymmetry
{\em independently of its shape}. 

\section{$AdS_4\times S^7$ and $AdS_7\times S^4$}
To find supersymmetric configurations in $AdS_4\times S^7$ we
embed $S^7$ into ${\bf C}^4$. Giant gravitons are
intersections of $S^7$ with the holomorphic surfaces in ${\bf C}^4$.
Spherical giant gravitons of \cite{MGST}
preserve $1\over 2$ of the supersymmetry. 
Gravitons of the form $F(Z_1,Z_2)=0$ preserve $1\over 4$ and those
given by $F(Z_1,Z_2,Z_3)=0$ as well as $F(Z_1,Z_2,Z_3,Z_4)=0$ preserve
$1\over 8$. The representation theory of $Osp(8|4,{\bf R})$ admits
also BPS multiplets preserving $3\over 8$ of the supersymmetry
\cite{FS}. But we do not expect to find the $3\over 8$ BPS
states in the spectrum of the superconformal theory on the
boundary of $AdS_4\times S^7$ \cite{SI}.

In $AdS_7\times S^4$ we  should find $1\over 2$ and $1\over 4$
BPS configurations \cite{FS}.
We  construct the corresponding giant gravitons in the following
way. Embed $S^4\subset {\bf R}^5$, and represent 
${\bf R}^5={\bf C}^2\times {\bf R}$. Consider  three-dimensional cylinders
in ${\bf C}^2\times {\bf R}$ of the form $C\times{\bf R}$ where
$C\subset {\bf C}^2$ is a holomorphic curve. Let us denote $Z_1$ and
$Z_2$ the coordinates in ${\bf C}^2$. The one-parametric group
of symmetries $Z_I\to e^{i\phi}Z_I$ preserves the $S^4$. Let us 
denote $e^{||}$ the corresponding vector field on $S^4$. 
The normalization of $e^{||}$ is such that $||e^{||}||^2=1$ on
the equator of $S^4$.
Consider $\Sigma=(C\times {\bf R})\cap S^4$ --- 
the intersection of the tube $C\times {\bf R}$ with $S^4$. 
Let us think of $\Sigma$ as the worldvolume of the
$M2$ brane at time zero, and let us move it with the speed of light 
along $e^{||}$. The corresponding brane configuration is supersymmetric.

Let us show that it is supersymmetric.
We denote $e^{\phi}$ the component of $e^{||}$ orthogonal to 
$T\Sigma$ normalized to $||e^{\phi}||^2=1$ and $\Gamma^{\phi}$
the gamma-matrix representation of $e^{\phi}$. 
The perpendicular component of the velocity is of the magnitude
$|\cos \nu|=|(e^{||}\cdot e^{\phi})|$.
The supersymmetry condition has a form very similar to
(\ref{SUSYForDThree}):
\begin{equation}\label{SUSYForMTwo}
(1+\Gamma^4\Gamma^{12})\left[\Gamma^{\tau}\Gamma^4\Gamma^n\Gamma^{\phi}
+\cos\nu \Gamma^n\Gamma^{12}-\sin\nu\right]\Psi_+=0
\end{equation}
Here $\Psi_+$ is a constant Majorana-Weyl spinor in ${\bf R}^{2+10}$ 
satisfying the following constraint:
\begin{equation}
\Gamma^5\cdots\Gamma^{12}\Psi_+=\Psi_+
\end{equation}
Let us denote $I$ the operator in $T{\bf R}^5=T({\bf C}^2\times{\bf R})$, 
which acts as zero on the vectors parallel to the ${\bf R}$-direction,
and as a complex structure on the vectors parallel to ${\bf C}^2$.
For any vector $v$, $I.v$ is parallel to ${\bf C}^2$.
Notice that $e^{\phi}$ is orthogonal to the ${\bf R}$-direction
in ${\bf C}^2\times {\bf R}$. (Let us explain why $e^{\phi}$ is
orthogonal to ${\bf R}$. Since $e^{||}$ is orthogonal
to ${\bf R}$, it is enough to show that the projection of
$e^{||}$ on $\Sigma$ is orthogonal to ${\bf R}$. Let us denote
$e^{\chi}$ the unit vector in $T\Sigma\cap {\bf R}^{\perp}$. Let us
show that the projection of $e^{||}$ on $T\Sigma$ is in fact
proportional to $e^{\chi}$. Indeed, $T\Sigma$ is generated by 
$e^{\chi}$ and the vector orthogonal to $e^{\chi}$. The vector
orthogonal to $e^{\chi}$ is a linear combination of $I.e^{\chi}$
and the vector parallel to ${\bf R}$. But $e^{||}$ is orthogonal
to $I.e^{\chi}$ because $I.e^{||}$ is orthogonal to $e^{\chi}$.
Therefore, $e^{||}$ is orthogonal to the generator of $T\Sigma$
perpendicular to $e^{\chi}$, and thus the projection of $e^{||}$
on $T\Sigma$ is parallel to $e^{\chi}$.)
In other words, both $e^{\phi}$
and $I.e^{\phi}$ are parallel to ${\bf C}^2$. Now let us impose
the following conditions on $\Psi_+$:
\begin{equation}
\Gamma^{\tau}\Gamma^4\Gamma(v)\Gamma(I.v)\Psi_+=\Psi_+\;\;
\mbox{for any}
\;\; v\in T{\bf C}^2\subset 
T({\bf C}^2\times{\bf R}),\;\; ||v||^2=1
\end{equation}
Using this condition in (\ref{SUSYForMTwo}) we get
\begin{equation}\label{AfterSubstitution}
(1+\Gamma^4\Gamma^{12})
\left[\Gamma^n\Gamma(I.e^{\phi})
+\cos\nu\Gamma^n\Gamma^{12}-\sin\nu\right]\Psi_+=0
\end{equation}
Notice that there is a relation of the form
\begin{equation}\label{LinearDependenceForMTwo}
I.e^{\phi}=\cos\alpha e^{\perp}+\sin\alpha e^n
\end{equation}
with some $\alpha$. Here $e^{\perp}$ denotes a vector orthogonal 
to $S^4$ in ${\bf R}^5$ (in our notations it is the same as $e^{12}$)
and $e^n$ denotes a vector in $TS^4$ orthogonal to both $e^{\phi}$
and $T\Sigma$.  
Indeed, the projection of $I.e^{\phi}$ to
$S^4$ should be orthogonal to both $T\Sigma$ and $e^{\phi}$.
(Let us explain why it is orthogonal to $T\Sigma$. The tangent
space to $\Sigma$ is generated by $e^{\chi}$ and $I.e^{\chi}+w$
where $w$ is some vector in the ${\bf R}$-direction. 
We have to show that $I.e^{\phi}$ is orthogonal to  $e^{\chi}$
and $I.e^{\chi}$. First of all,
$(I.e^{\phi}\cdot e^{\chi})=c(I.e^{||}\cdot e^{\chi})=
0$ since $I.e^{||}$ is a sum of $e^{\perp}$ and a vector
in ${\bf R}$-direction. Also, $(I.e^{\phi}\cdot I.e^{\chi})=
(e^{\phi}\cdot e^{\chi})=0$. Therefore $I.e^{\phi}$ is orthogonal
to $T\Sigma$.) Therefore there is a relation of the form 
(\ref{LinearDependenceForMTwo}). This relation implies that
(\ref{AfterSubstitution}) is satisfied provided that $\nu=\alpha+\pi$.
(Remember that in our notations $\Gamma^{12}$ is the same as
$\Gamma(e^{\perp})$.) We need more work to verify that
$\nu=\alpha+\pi$.

Let us denote $h$ the coordinate on ${\bf R}$, so that the equator
of $S^4$ is at $h=0$ and the poles of $S^4$ are at $h=\pm 1$.
For $|h|<1$ we introduce $\theta$, $h=\cos\theta$.
Let us denote $E^{||}={1\over\sin\theta}e^{||}$. We have
$||E^{||}||^2=1$.
For any $h_0\in {\bf R}$ we can consider the plane $H_h$ consisting
of the points with $h=h_0$. We denote $S_h^3=H_h\cap S^4$ and
$E^{\perp}$ the unit vector normal to $S_h^3$ in $H_h$.
Now $E^{||}$ can be decomposed as follows:
\begin{equation}\label{RelationForE}
E^{||}=-\cos\beta e^{\phi} +\sin\beta e^{\chi}
\end{equation}
(This is just a statement that the projection of $e^{||}$ on 
$T\Sigma$ is proportional to $e^{\chi}$ and the orthogonal projection
is proportional to $e^{\phi}$.)
This implies that
\begin{equation}\label{CosNu}
\cos\nu=-\sin\theta\cos\beta
\end{equation}
Let us consider the intersection $\gamma=\Sigma\cap S_h^3$. 
This $\gamma$ is a string (real one-dimensional curve) in $S_h^3$. 
The trajectory of this string in $S_h^3$ is exactly the same 
as the motion of the D-string in the $S^3$ of $AdS_3\times S^3$
considered in Section 4. Let $E^n$ be the unit vector in $TS_h^3$
orthogonal to both $T\gamma$ and $e^{\phi}$.  
The analysis similar to Section 3.2 shows that
\begin{equation}\label{LinearDependenceInPlane}
I.e^{\phi}=\cos\beta E^{\perp} \pm \sin\beta E^n
\end{equation}
where $\beta$ is the same $\beta$ as in (\ref{RelationForE}).
Let us compare this relation with (\ref{LinearDependenceForMTwo}).
Taking a scalar product of (\ref{LinearDependenceForMTwo}) 
and (\ref{LinearDependenceInPlane}) with $e^{\perp}$ we get:
\begin{equation}\label{CosAlpha}
\cos\alpha=(e^{\perp}\cdot E^{\perp})\cos\beta=\sin\theta\cos\beta
\end{equation}
Now (\ref{CosNu}) and (\ref{CosAlpha}) imply that $\nu=\pm\alpha+\pi$
up to sign. This is what we needed (the sign 
can be adjusted by changing the orientation of $M2$).

\section{Energy and Angular Momentum}
In this section we consider giant gravitons in $AdS_5\times S^5$.
\subsection{Energy of the Giant Graviton}
The energy is given by the following formula:
\begin{equation}
E=\int_{\Sigma}d^3\sigma{1\over \sqrt{1-v_{\perp}^2}}=
\int_{\Sigma}d^3\sigma{1\over |(e^{||}\cdot e^{\psi})|}=
\int_{\Sigma}d^3\sigma{1\over |(e^{\perp}\cdot I.e^{\psi})|}
\end{equation}
Notice that $(e^{\perp}\cdot I.e^{\psi})$ is the cosine of the angle
between $C$ and the normal vector to $S^5$. This enables us to express
the integral as follows:
\begin{equation}
E=\int_C d\;\mbox{vol}(C)\;\delta\left(\sqrt{|Z_1|^2+|Z_2|^2+|Z_3|^2}-1
\right)
\end{equation}
This is the mass formula.
\subsection{BPS bound}
Here we will express the energy of the giant graviton in terms
of its momentum. Let us remember the definition
of the momentum. Consider a classical mechanical system with the action
being invariant under the global symmetry. This means that when we vary
the trajectory $\delta \phi^i(t)=\epsilon \xi^i(\phi(t))$ with $\xi^i$
being the generator of the symmetry, the action does not change:
\begin{equation}
\delta_{\epsilon}S=
\epsilon \int dt \xi^i(\phi(t)){\delta S\over\delta \phi^i(t)}=0
\end{equation}
Suppose now that $\epsilon$ depends on $t$. Then, the variation of
the action is:
\begin{equation}
\delta_{\epsilon(t)}S=\int dt \dot{\epsilon}(t)M(\phi(t))
\end{equation}
where $M(\phi)$ is the momentum. On the classical trajectories,
the variation of the action is zero anyway;
this implies that $\dot{M}(\phi(t))=0$ --- the momentum conservation.

In our situation the action is invariant under $SO(6)$.
Consider a subgroup $U(1)_I\subset SO(6)$ generated by
the overall phase rotations $Z_I\to e^{it}Z_I$, $I=1,2,3$.
Let $M$ denote the corresponding momentum. 
The corresponding infinitesimal symmetry is 
$\delta Z_I=i\epsilon Z_I$. Let us compute $M$. 
The Born-Infeld action consists of two terms, 
$\int\sqrt{\mbox{det}\partial X\cdot \partial X}$ and 
the Wess-Zumino term $S_{WZ}$. We will denote these contributions
$M_0$ and $M'$, respectively. Let us start with calculating $M_0$:
\begin{equation}
M_0=\int_{\Sigma}d\sigma {v_{\perp}^2\over \sqrt{1-v_{\perp}^2}}
=\int_{\Sigma}d\sigma \left( {1\over (e^{||}\cdot e^{\psi})} 
-(e^{||}\cdot e^{\psi})\right)
\end{equation}
This looks like energy, except for the second term, 
$-\int_{\Sigma}d^3\sigma (e^{||}\cdot e^{\psi})$.
We want to show that this term cancels by the contribution from the
Wess-Zumino term.  
Indeed, the contribution of the 
Wess-Zumino term to the angular momentum can be written as follows:
\begin{equation}\label{DeltaM}
\Delta M=\int_{\partial^{-1}\Sigma}\iota (e^{||})F
\end{equation}
where $\partial^{-1}\Sigma$ is a surface with the boundary $\Sigma$,
and $F$ is the five-form, which is proportional
 to the volume form\footnote{The coefficient of proportionality 
for $AdS_5\times S^5$ is 4. One can remember it in the following way.
Consider the spherical $D3$ brane in $AdS_5$ approaching the boundary
at $r=\infty$.
The leading contributions to the action from the Born-Infeld
term and the action should cancel when $r\to\infty$ \cite{SW}.
Suppose that we change $r$ a little bit. The Born-Infeld part of
the action will change as $\int dt \delta(r^4 \mbox{vol}(S^3))$ 
(we are using the metric $-(1+r^2)dt^2+{dr^2\over 1+r^2}
+r^2ds^2_{S^3}$ on $AdS_5$). The change in the Wess-Zumino
coupling will be $C\int dt \delta r r^3 \mbox{vol}(S^3)$ where $C$
is the coefficient of proportionality between $F_5$ and the volume
form. Therefore $C=4$.} on $S^5$. 
One can write 
$F=4\iota(e^{\perp}).{1\over 6}\omega\wedge\omega\wedge\omega$
where $\omega$ is a Kahler form on ${\bf C}^3$. Also, 
notice that for the restriction on the tangent space to $S^5$,
we have:
\begin{equation}
(\iota(e^{||}).\iota(e^{\perp}).\omega\wedge\omega\wedge\omega)|_{TS^5}=
3(\omega\wedge\omega)|_{TS^5}
\end{equation}
(this is because $\iota(e^{||}).\omega|_{TS^5}=0$.)
Let us call $C_+$ the part of $C$ which is inside $S^5$
and $C_-$ the part of $C$ which is outside $S^5$.
Using $d(\omega\wedge\omega)=0$ we can rewrite (\ref{DeltaM}) as follows:
\begin{equation}
\Delta M=2\int_{C_+} \omega\wedge\omega
\end{equation}
Now let us use the fact that 
$$
\omega={1\over 4}d\sum\limits_{I=1}^3(\overline{Z}^IdZ^I-
Z^I d\overline{Z}^I)
$$
This allows us to write
\begin{equation}
\Delta M={1\over 2}\int_{\partial C_+=\Sigma} \omega\wedge
\sum\limits_{I=1}^3(\overline{Z}^IdZ^I-
Z^I d\overline{Z}^I)
\end{equation}
One can see that $\omega|_{\Sigma}$ is the area for on $T_0\Sigma$
and
$$\sum\limits_{I=1}^3(\overline{Z}^IdZ^I-
Z^I d\overline{Z}^I)=
2(e^{\perp}\cdot I.e^{\psi})d\psi$$
where $d\psi$ is the length form on the orthogonal complement
to $T_0\Sigma$ in $T\Sigma$. Therefore, 
\begin{equation}
\Delta M= \int_{\Sigma}d^3\sigma (e^{\perp}\cdot I.e^{\psi})
=\int_{\Sigma}d^3\sigma (e^{||}\cdot e^{\psi})
\end{equation}
which is what we wanted to prove. We see that the energy of the giant
graviton is equal to the value of the angular momentum on the
generator of $U(1)_I\subset SO(6)$.
This is the BPS bound.


\subsection{Some examples}
We have not computed the angular momentum of the giant graviton in the
closed form. However, for some giant gravitons we can predict
the {\em direction} of the angular momentum. Let us consider the giant
graviton corresponding to the following surface $C$:
\begin{equation}
Z_1^{a_1}Z_2^{a_2}Z_3^{a_3}=\rho
\end{equation}
The worldvolume of this giant graviton is invariant under
$U(1)^3\subset SO(6)$. The first $U(1)$, which we will call $U(1)_I$,
is just $Z_I\to e^{it}Z_I$. The second and the third $U(1)$, which
we will call $U(1)^2_{(a_1,a_2,a_3)}$, is $Z_I\to e^{isw_I}Z_I$
where $\sum\limits_{I=1}^3 w_Ia_I=0$. These three $U(1)$ generate
the maximal torus of $SO(6)$. Since the worldvolume is invariant
under $U(1)^3$, the angular momentum should commute with $U(1)^3$.
Therefore it belongs to the Cartan subalgebra.
One can also see that the angular momentum is orthogonal to the
generators of $U(1)^2_{(a_1,a_2,a_3)}$. 
In our situation, generators of $U(1)^2_{(a_1,a_2,a_3)}$ act as 
reparametrizations on the worldvolume. Since the Born-Infeld action
for the $D3$ brane is reparametrization invariant, it follows that
$\delta_{\epsilon}S$ is zero independently of whether or not 
$\epsilon$ depends on $t$. Therefore, the component of the angular
momentum corresponding to $U(1)^2_{(a_1,a_2,a_3)}$ is zero.
In fact, there is a subtlety here. The Born-Infeld action includes
the Wess-Zumino term, and we have to exercise care in 
defining it\footnote{I want to thank O.~DeWolfe for discussions on the
role of the Wess-Zumino term.}.
Given the worldsurface $\Sigma(t)$, let us consider the four-dimensional
surface $\Lambda(t)$, such that its boundary is $\Sigma$: 
$$\partial\Lambda(t)=\Sigma(t)$$
The Wess-Zumino term in the action is the integral of
the five-form field strength over the surface swept by $\Lambda(t)$.
The contribution of the Wess-Zumino term to the angular momentum 
is zero provided that the symmetry acts as a reparametrization
not only on $\Sigma$, but also on $\Lambda$. In our case, we can 
choose $\Lambda$ to be a family of surfaces starting with $\Sigma$,
parametrized by $k$, $1\leq k<\infty$. This family is described
by the equation:
\begin{equation}
Z_1^{a_1}Z_2^{a_2}Z_3^{a_3}=k\rho,\;\;\; 1\leq k<\infty
\end{equation}
One can see that $U(1)^2_{(a_1,a_2,a_3)}$ preserves this family,
therefore the contribution of the Wess-Zumino term 
to the $U(1)^2_{(a_1,a_2,a_3)}$ component of the momentum is zero.
The angular momentum of our graviton is a linear combination
of the generators of $U(1)_I$ and $U(1)^2_{(a_1,a_2,a_3)}$, orthogonal 
to the generators of $U(1)^2_{(a_1,a_2,a_3)}$. Therefore, it has the 
following form:
\begin{equation}
M= 
\mu\left( \begin{array}{cccccc}
0&-a_1&0&0&0&0\\
 a_1&0&0&0&0&0\\
0&0&0&-a_2&0&0\\
 0&0&a_2&0&0&0\\
0&0&0&0&0&-a_3\\
 0&0&0&0&a_3&0
\end{array}
\right)
\end{equation}
Here we have used the basis $(\mbox{Re}Z_1,\mbox{Im}Z_1,
\mbox{Re}Z_2,\mbox{Im}Z_2,\mbox{Re}Z_3,\mbox{Im}Z_3)$ in ${\bf R}^6$.
The coefficient $\mu$ is a function of $\rho$.

\section{Supersymmetric branes in flat space}
It is interesting that the analogue of our construction works in
flat space. The proof is actually simpler in flat space.
In flat space we can get only noncompact BPS configurations
(extending to the spatial infinity).
We consider Type IIB in a flat ten-dimensional Minkowski space.
\subsection{$D1$ brane}
Here we want to consider a well known example of a supersymmetric
moving brane in Type IIB
which is in some sense analogous to those configurations which we 
have constructed in AdS times sphere.
The $D1$ brane {\em of arbitrary shape} moving with the speed of light 
in arbitrary direction
$e^{||}$ preserves $1\over 4$ of the supersymmetry\footnote{Suppose
that at infinity the brane approaches a straight line parallel 
to $e^{||}$; in this case our configuration would be called ``left-moving
wave''; it preserves $1\over 2$ SUSY on the worldvolume theory
of a D-string, and correspondingly $1\over 4$ SUSY
of the target space theory.}.
The preserved generators satisfy 
\begin{equation}\label{ConditionForDOne}
\Gamma^0\Gamma (e^{||})\psi=\psi,\;\;\; \psi=i\psi^*
\end{equation}
Let us prove it. Suppose that 
\begin{equation}\label{LinearForDOne}
e^{||}=-\cos\alpha e^{\phi}+
\sin\alpha e^{\psi}
\end{equation}
where $e^{\phi}$ is orthogonal to 
the worldline of the $D1$ brane and $e^{\psi}$ belongs to
the worldline. The supersymmetry preserved in the presence
of a D string is $\Gamma^0\Gamma^1\psi=i\psi^*$.
Using $\psi=i\psi^*$ we can rewrite this condition as follows:
\begin{equation}\label{SUSYForDOne}
{\Gamma^0-\cos\alpha\Gamma(e^{\phi}) \over\sin\alpha}
\Gamma(e^{\psi})
\psi=\psi
\end{equation}
One can see that (\ref{ConditionForDOne}) and
(\ref{LinearForDOne}) implies (\ref{SUSYForDOne}).

\subsection{$D3$ brane}
Some examples of supersymmetric $D3$ branes in flat space
can be obtained from the supersymmetric $D1$ branes of the previous
subsection. Their worldvolume is a product of a real one-dimensional
curve $\gamma$ a complex curve $\beta$ which is orthogonal to $\gamma$.
To preserve supersymmetry this brane should move with the speed of light 
in the direction orthogonal to $\beta$.

There are more general examples.
Consider the five-dimensional plane $M\subset {\bf R}^{1+9}$.
We will embed $M\subset {\bf C}^3$. Again, we have
$e^{||}$ --- the preferred direction in $TM$, $e^{||}=Ie^{\perp}$.
Let $C$ be the complex surface in ${\bf C}^3$, $\Sigma=C\cap M$,
and move $\Sigma$ with the speed of light in the direction $e^{||}$.
We denote $\cos\alpha e^{\phi}$ the projection of $e^{||}$ to the
direction orthogonal to $T\Sigma$, 
and it turns out that $e^{\phi}$ is the unit vector in 
the intersection of $(TC)^{\perp}$ with $TM$, and the tangential 
component is proportional to $e^{\psi}$, just as in 
(\ref{ExpressionForPrefferedE}). 

The supersymmetry condition:
\begin{equation}
\left[i(\Gamma^0+\cos\alpha \Gamma(e^{\phi}))\hat{\Sigma}+
\sin\alpha\right]\psi=0
\end{equation}
Let us impose the following conditions on $\psi$:
\begin{equation}
\begin{array}{l}
1)\;\Gamma^0\Gamma(e^{||})\psi=\psi,\\
2)\;\gamma(v)\gamma(Iv)\psi=i||v||^2\psi, \;\;
\mbox{for any}\;v\in T_0M
\end{array}
\end{equation}
(we denote $T_0M$ the subspace of $TM$ orthogonal to $e^{||}$).
Notice that from the second condition follows
$\hat{\Sigma}\psi=i\Gamma(e^{\psi})\psi$. Now we can rewrite the 
supersymmetry condition as follows:
\begin{equation}
\begin{array}{c}
0=\left[-(\Gamma^0+\cos\alpha\Gamma^{\phi})\Gamma(e^{\psi})+
\sin\alpha\right]\psi
=\\=
\left[\Gamma(e^{\psi})(\Gamma^0+\cos\alpha\Gamma(e^{\phi}))+
\sin\alpha\right]\psi
=\\=
\left[\Gamma(e^{\psi})(\Gamma(e^{||})+\cos\alpha\Gamma(e^{\phi}))
+\sin\alpha\right]\psi
\end{array}
\end{equation}
which is the consequence of the first of Eqs. 
(\ref{ExpressionForPrefferedE}).

\appendix
\section{Giant gravitons in perturbation theory}
The authors of \cite{DJM} have studied excitations around the
spherical giant gravitons. They have used the quadratic approximation 
to the action and computed the frequencies of the normal modes.
It was found that the excitations along AdS decouple in the
quadratic approximation from the excitations along the sphere.
Let us look at those excitations which are inside the sphere,
which means $v_k=0$ in the notations of \cite{DJM}. 
Suppose that the sphere has dimension $n$. The frequencies 
are given by the formula (4.17 of \cite{DJM}):
\begin{equation}
\omega_{\pm}^2=Q+{(n-3)^2\over 2}\pm (n-3)
\sqrt{Q+{(n-3)^2\over 4}}
\end{equation}
where $Q$ is the eigenvalue of the Laplace operator on a sphere 
of dimension $n-2$ with the radius $1$. The possible values for
$Q$ are $N(N+n-3)$ where $N$ is an integer. Therefore
$\omega_+=\pm(N+n-3)$ and $\omega_-=\pm N$. The frequencies are integer,
therefore the motion is periodic. The configuration returns to
itself after its projection on AdS travels along the geodesic the distance 
equal to the  circumference of the sphere.  
But this is exactly what we expect to see when we study in perturbation
theory the $1\over 4$ or $1\over 8$ supersymmetric giant gravitons 
which are close to the given spherical configuration.
Indeed, the flow generated by the vector field $e_{||}$ is periodic with
the same period. 

Most of the excitations found in \cite{DJM} are not BPS. But some
of them should be BPS and should represent our configurations 
in the linearized approximation.

\vspace{25pt}
\hspace{-21.5pt}
{\Large \bf Acknowledgements}\vspace{12pt}\\
I would like to thank 
O.~DeWolfe, D.~Gross, A.~Hashimoto, S.~Hirano, N.~Itzhaki and E.~Witten 
for illuminating discussions. 
This work was supported in part by the NSF Grant No. PHY99-07949, 
and in part by RFFI Grant No. 00-02-16477 and by the
Russian grant for the support of the scientific schools No. 00-15-96557.

\end{document}